\documentclass[reprint, amsmath, amssymb, amsfonts, showpacs, superscriptaddress, floatfix, aps, pra]{revtex4-1}
\usepackage{subfigure}
\usepackage{graphicx}
\usepackage{braket}
\usepackage[breaklinks=true,colorlinks=true,linkcolor=blue,urlcolor=blue,citecolor=blue]{hyperref}
\newcommand{\expect}[1]{\mathinner{\langle #1\rangle}}
\begin{document}

\title{Nonreversal and nonrepeating quantum walks}

\author{T. J. Proctor}
\thanks{The first two authors contributed equally to this work}
\author{K. E. Barr}
\thanks{The first two authors contributed equally to this work}
\author{B. Hanson}
\affiliation{School of Physics and Astronomy, E. C. Stoner Building, University of Leeds, Leeds 
LS2 9 JT}
\author{S. Martiel}
\affiliation{School of Physics and Astronomy, E. C. Stoner Building, University of Leeds, Leeds 
LS2 9 JT}
\affiliation{Universit\`{e} Nice Sophia Antipolis, Laboratoire I3S, UMR 7271, 2000 Route des Colles, 06903 Sophia Antipolis, France}
\author{V. Pavlovi\'{c}}
\affiliation{School of Physics and Astronomy, E. C. Stoner Building, University of Leeds, Leeds 
LS2 9 JT}
\affiliation{Faculty of Science and Mathematics, University of Nis, Serbia}
\author{A. Bullivant}
\author{V. M. Kendon}
\affiliation{School of Physics and Astronomy, E. C. Stoner Building, University of Leeds, Leeds 
LS2 9 JT}

\date{\today}

\begin{abstract}
We introduce a variation of the discrete time quantum walk, the nonreversal quantum walk, which does not step back onto a position which it has just occupied. This allows us to simulate a dimer and we achieve it by introducing a new type of coin operator. The nonrepeating walk, which never moves in the same direction in consecutive time steps, arises by a permutation of this coin operator. We describe the basic properties of both walks and prove that the even-order joint moments of the nonrepeating walker are independent of the initial condition, being determined by five parameters derived from the coin instead. Numerical evidence suggests that the same is the case for the nonreversal walk. This contrasts strongly with previously studied coins, such as the Grover operator, where the initial condition can be used to control the standard deviation of the walker. 
\end{abstract}


\maketitle

\section{Introduction}
Quantum walks have been extensively studied since their introduction \cite{gudder1988quantum, PhysRevA.48.1687,  grossing1988quantum}. Initial interest in cellular automata \cite{grossing1988quantum} led to more general algorithmic applications \cite{farhi1998quantum,AAKV,Ambainis01} precipitating a rapid development of the theory of computation by quantum walks. 
Quantum walks can provide quadratically enhanced searching \cite{shenvi2003quantum,magniez09a}
with generalizations to related computational tasks such as element distinctness
\cite{Ambainis04} and subset finding \cite{childs03b,magniez05a}. 
Quantum walks have been shown to have interesting transport properties in a variety of scenarios. On the line they achieve ballistic transport \cite{Ambainis01} and they were first shown to have an exponential speedup over the classical random walk on the hypercube by Kempe \cite{kempe02a,Kempe05} for the discrete time walk  and Childs et al.~for the continuous time walk \cite{Childs02}, followed by an algorithm with a proven exponential speed up \cite{Childs03}. 

\begin{figure}[h!]
\rotatebox{0}{\includegraphics[scale = 0.25]{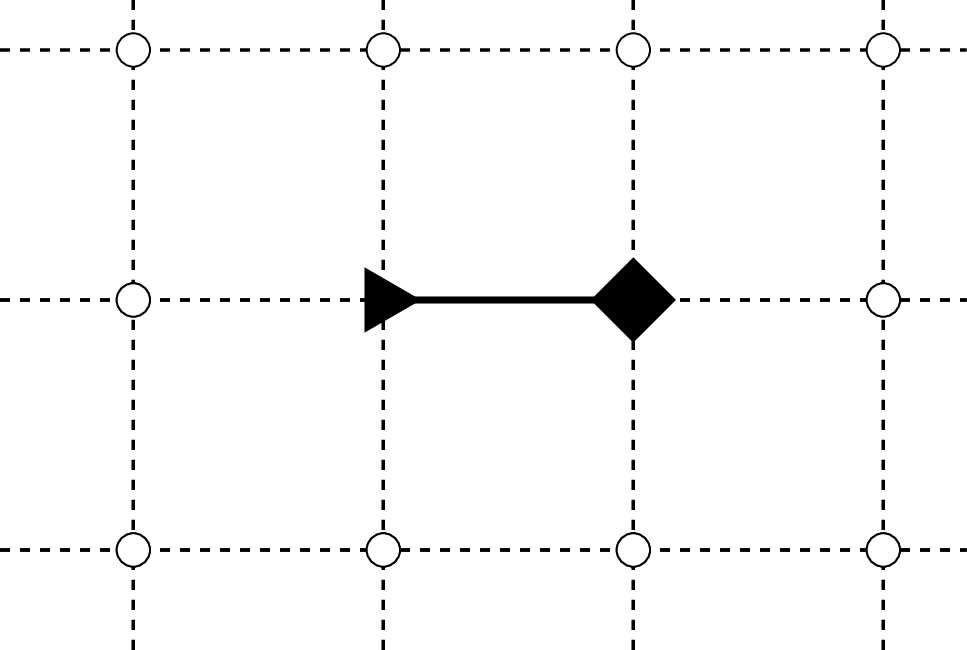}} 
\caption{The head and tail of a dimer on a square lattice, the diamond marking the head, and the triangle marking the tail. 
As the tail prevents the head moving left, the head can move up, down, or right, resulting in a nonreversal walk} 
\label{fig:dimer}
\end{figure}

The quantum walks introduced thus far model idealized walkers with no spatial extension. Whilst these have many uses in modeling physical and biological processes, e.g., \cite{Mohseni08}, we may also want to consider walkers which do have spatial extension, and hence can only move into positions which they are not already occupying. In a classical setting, self-avoiding random walks were developed to model precisely such processes, initially the folding of polymers. The simplest case of self-avoidance is a dimer occupying two adjacent lattice sites.  For a dimer with distinguishable halves, a ``head'' followed by a ``tail'', self-avoidance means the head cannot step back onto the previously occupied position, since that is now occupied by the tail, see Fig.~\ref{fig:dimer}.  This is thus known as the nonreversal walk. In this paper, we introduce a quantum version of such a walk. The motivation for studying the nonreversal quantum walk is much the same as that for studying the classical version: more realistic simulation of physical systems.

In both the classical and quantum cases, the self-avoiding or nonreversal walk on the line is trivial. This is because there are only two degrees of freedom in the movement, so if one of those is prohibited by the model, then unidirectional ballistic transport is obtained. The walks studied in this paper take place over a square lattice, in which case the dynamics are highly non-trivial. The paper proceeds as follows: in Section \ref{sec:classical} classical self-avoiding and nonreversal walks are described in more detail, to provide background and context. Then for the sake of comparison, the properties of the quantum walk on the square lattice are briefly outlined in Section \ref{sec:2d}. The nonreversal and the closely related nonrepeating quantum walks are then defined in Section \ref{sec:analytics}. The properties of the nonrepeating walk are  explored analytically in Section \ref{sec:nr}. With the aid of numerical simulations, the nonreversal walk is investigated in Section \ref{sec:numerics}. We finish in Section \ref{sec:conclusion} with some concluding remarks.

\subsection{Classical self-avoiding random walks}
\label{sec:classical}
The classical self-avoiding walk has proven difficult to treat analytically, hence the results concerning it have all so far been numerical \cite{Hayes98} and there remain many open questions. Even enumerating the number of self-avoiding walks has proven very difficult, despite them being so rare that coming upon one by mistake when examining a random walk is highly improbable. If we denote by $c_n$ the number of self-avoiding walks of precisely $n$ steps, then the total number of self-avoiding walks up to length $n$ is $\sum_{n \geq 2} c_n$. Some facts are clear, for example that $c_{n+m} \leq c_n + c_m$. The set of self-avoiding walks of length $n$ concatenated with those of length $m$ contains not only the self-avoiding walks of length $n+m$ but some which overlap, hence the inequality. While determining the precise number of walks is difficult, some bounds have been established. On a square lattice, the number of nonreversal walks of length exactly $n$ steps is $3^n$, since there are three choices of direction at each step.  The number of self-avoiding walks must be less than $3^n$, as the nonreversal walks include the self-avoiding walks as a subset. Additionally, it is possible to construct subsets of self-avoiding walks which grow as $2^n$, hence we know that there are between $2^n$ and $3^n$ self-avoiding random walks. The best evidence so far suggests that the number of self-avoiding walks of length $n$ is proportional to $2.638^n$, and this is provided as a non-rigorous estimate in \cite{Jensen03}. The evidence for this value was obtained by enumerating each such walk of length up to 51 and required a 1024 processor supercomputer \cite{Guttmann01}. Without new algorithms it is unlikely that we will be able to enumerate much further than this. 

As even counting the walks has proved difficult, it is unsurprising that little is known regarding other properties. Interesting quantities with which to compare different walks include the average distance from the origin, denoted $\expect{r}$, the average of the square of this distance $\expect{r^2}$, and the standard deviation of $r$. 
For the self-avoiding walk, $\expect{r^2}$ is conjectured to be proportional to $n^{3/2}$ though so far, even a proof that the exponent must be between 1 and 2 is elusive \cite{Hayes98}. Another interesting property of self-avoiding walks demonstrates a key difference between the self-avoiding walk and its standard and nonreversal counterparts.  The self-avoiding walk does not necessarily continue to evolve indefinitely.  This is because it is possible to reach a lattice site whose only adjacent lattice sites have previously been visited, hence the walker becomes stuck.  

The nonreversal walk is in some ways more tractable.  As already noted, on the square lattice there are $3^n$ such walks of exactly $n$ steps.  Its mean squared displacement is $\expect{r^2}=2n$, so it spreads twice as fast as the standard random walk. There is very little literature on the nonreversal walk, and what there is tends to examine specific characteristics of the walk relevant to the study of polymer chains \cite{Skliros10}, rather than its general features.
 
\subsection{Quantum walks on the square lattice}
\label{sec:2d}
We first define the formalism for the discrete time quantum walk on a square lattice before discussing previous results for such quantum walks. The walk is defined on $\mathbb{Z}^2 = \{ (x,y) : x,y \in \mathbb{Z} \} $ where $\mathbb{Z}$ denotes the set of integers. The state of the system, $\Psi$, is then described by a four-dimensional vector at each lattice site, corresponding to four possible coin states that are internal degrees of freedom of the walker. We denote this as:
\begin{equation}   \Psi(x,y,t) = \begin{pmatrix} \psi^{x+}(x,y,t)  \\ \psi^{y+}(x,y,t) \\ \psi^{y-}(x,y,t) \\ \psi^{x-}(x,y,t) \end{pmatrix}  ,\end{equation}
where each component is a complex function of the discrete position of the walker, $(x,y)$, and discrete time $t$ and where $\sum_{x,y,j} |\psi^{j}(x,y,t)|^2=1$ with $j$ taking the symbols $x+,y+,y-,x-$. These four coin states are associated with the walker moving in the positive $x$, positive $y$, negative $y$ and negative $x$ directions respectively.

The evolution is then defined by a coin operator, which acts only on the coin subspace of the walker, and a shift operator which acts on the entire Hilbert space. The coin operator at a particular site is therefore an operator in $SU(4)$. Different coin operators can in general be chosen for different lattice sites and they may vary in (discrete) time. In what follows the same coin operator is chosen at all lattice sites and it does not vary in time. We denote the coin operator that acts on the state of the walker (and so is constructed from the individual coin operators at each site) as $C^{c}$, where $c$ labels a particular choice of coin operator. The shift operator, $S$, is defined by: 
\begin{equation} S \Psi(x,y,t) =  \begin{pmatrix} \psi^{x+}(x-1,y,t)  \\ \psi^{y+}(x,y-1,t) \\ \psi^{y-}(x,y+1,t)\\ \psi^{x-}(x+1,y,t) \end{pmatrix} \label{shift}.\end{equation} 
Therefore, the action of the shift operator is to move the $\psi^{x+}$ coin state at a particular vertex $(x,y)$ one step in the positive $x$ direction to the vertex $(x+1,y)$, and analogously for the three other coin states. This can be seen from the definition of the shift operator as the $\psi^{x+}$ coin state at $(x,y)$ depends on the pre-shift $\psi^{x+}$ coin state at $(x-1,y)$. We then define the operator that evolves the walk by one time step $U_c$, by the action of the coin operator $C^c$, followed by the shift operator, $S$. That is:
\begin{equation} \Psi(x,y,t+1) =  U_c \Psi(x,y,t) = S\cdot C^{c} \Psi(x,y,t). \end{equation}
The choice of coin operator and the initial state of the walker then completely define the walk as we may write
\begin{equation} \Psi(x,y,t) =  U_c^{t} \Psi(x,y,0) = \left(S\cdot C^{c} \right)^{t} \Psi(x,y,0). \end{equation}

\begin{figure}[th!]
  \subfigure 
    {(a)\resizebox{0.8\columnwidth}{!}{\includegraphics[trim= 2cm 1.4cm 2cm 1.4cm]{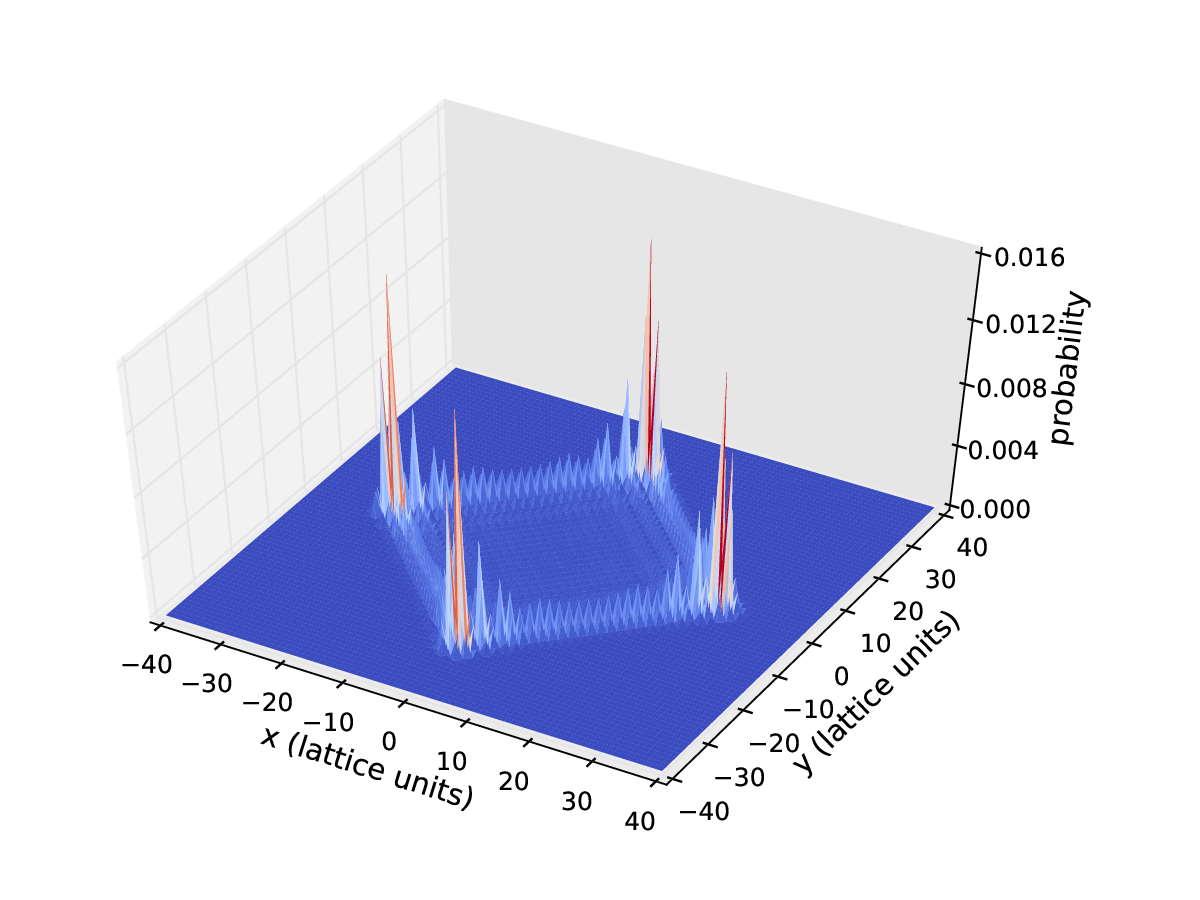}}} \\ 
  \subfigure
    {(b)\resizebox{0.8\columnwidth}{!}{\includegraphics[trim= 2cm 1.4cm 2cm 1.4cm]{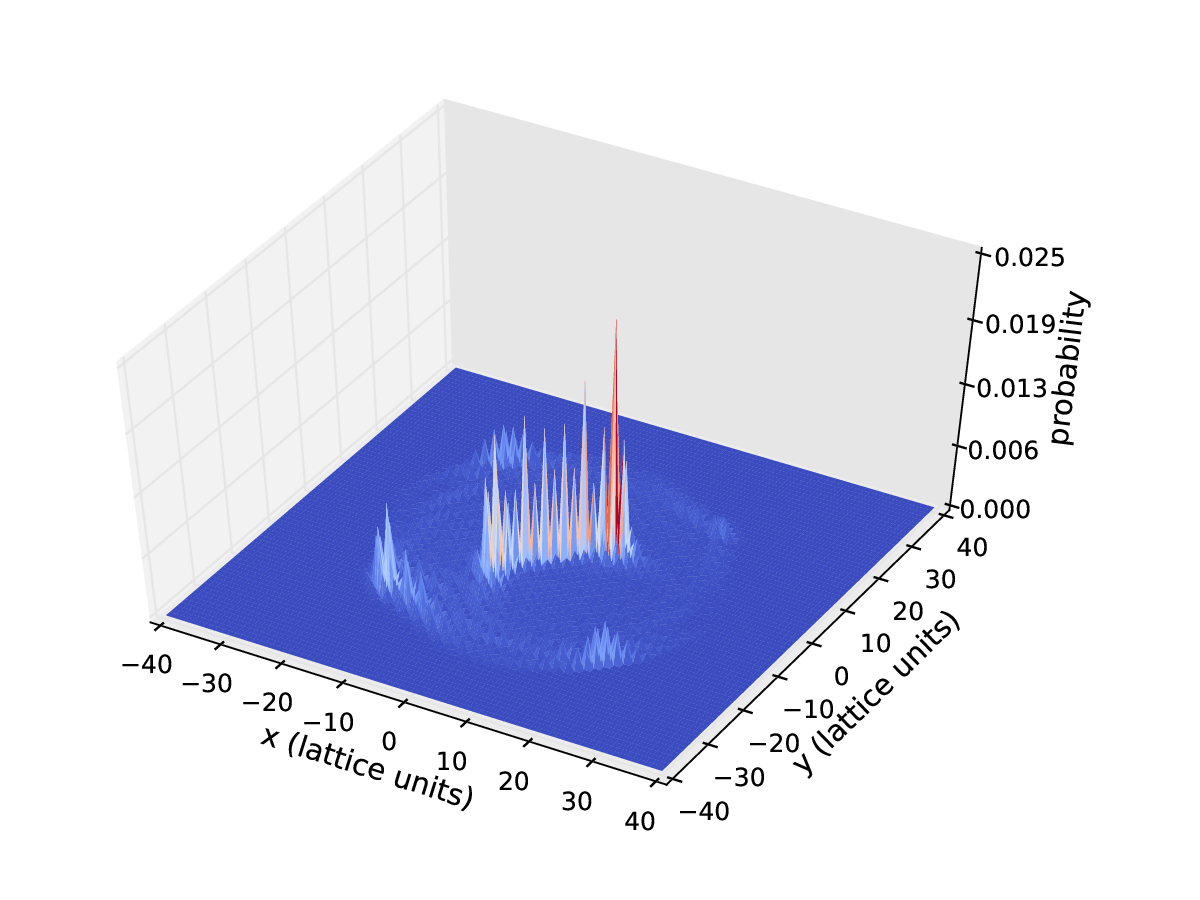}}} \\
  \subfigure 
    {(c)\resizebox{0.8\columnwidth}{!}{\includegraphics[trim= 2cm 1.4cm 2cm 1.4cm]{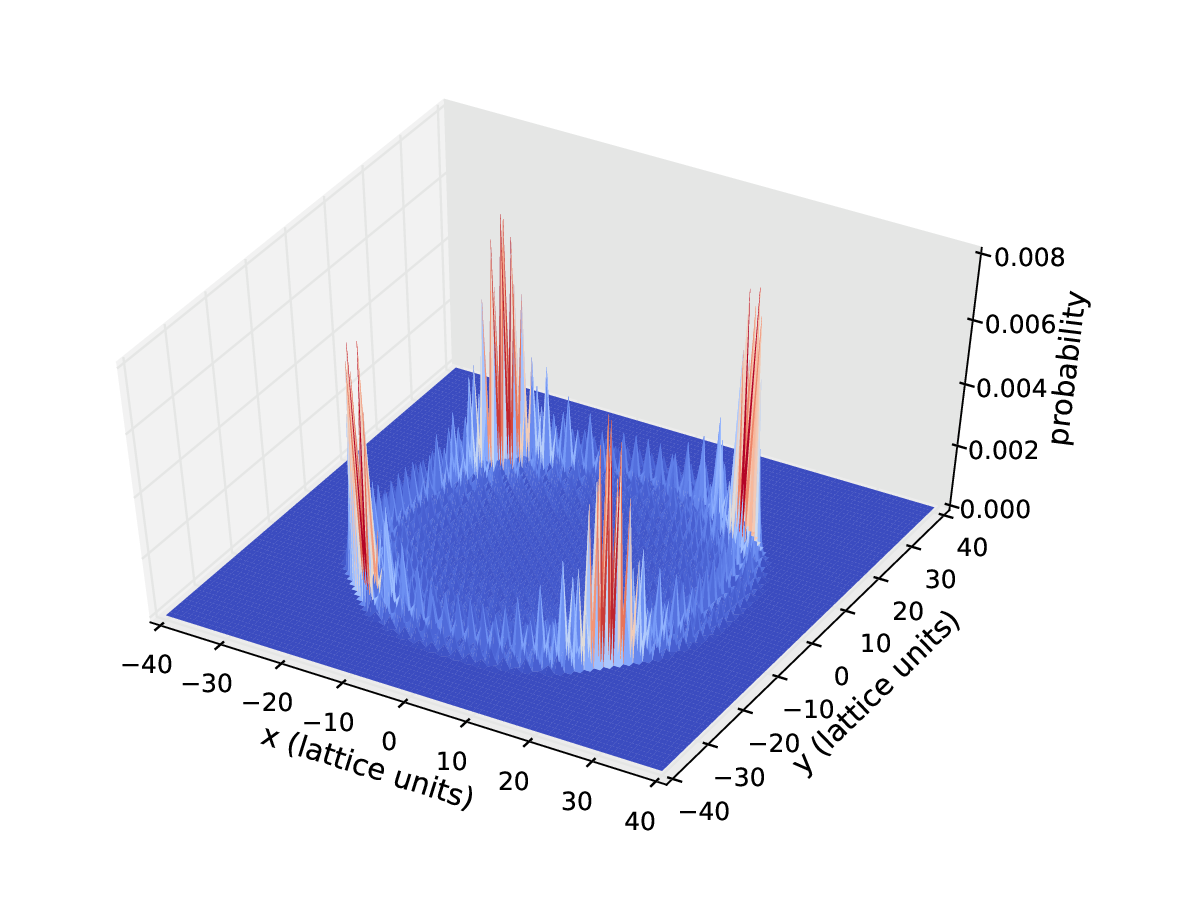}}}
\caption{(Color online) Probability distributions arising from (a) the Hadamard coin and b) the DFT coin for the initial state of Eq.~(\ref{eqn:initcond1}), and c) the Grover coin
for initial state of Eq.~(\ref{eqn:initcond2}).} 
\label{fig:2dwalks}
\end{figure}
The properties of the discrete time quantum walk on the square lattice were extensively explored in \cite{Tregenna03} following initial investigations in \cite{Mackay02}. In particular they examined the mean distance from the origin at time $t$:
\begin{equation}
\label{eqn:mean}
 \langle r \rangle_t = \sum_{x, y} p(x,y, t)\, \sqrt{x^2 + y^2} ,
\end{equation}
where $r$ is the radial distance from the origin, and $p(x,y, t)$ is the probability of finding the walker at position $(x,y)$ at time $t$. They also characterise the walks in terms of the standard deviation of $r$,
\begin{equation}
\label{eqn:sd}
\sigma = \sqrt{\langle r^2 \rangle - \langle r \rangle ^2 },
\end{equation}
which characterises how spread out the walker is over the lattice.  Larger $\sigma$ indicates greater spreading or variation in $r$, while small $\sigma$ indicates the walker is moving out radially in a well-defined ring.
The authors of \cite{Tregenna03,Mackay02} carried out a comparison between three different coin operators. Their first choice is a tensor product of two Hadamard operators for the walk on a line:
\begin{equation}
H \otimes H  = \frac{1}{2}\begin{pmatrix} 1 & 1 & 1 & 1 \\ 1 & -1 & 1 & -1 \\ 1 & 1 & -1 & -1\\ 1 & -1 & -1 & 1\\ \end{pmatrix}. \label{Hadcoin}
\end{equation}
 This creates a separable unitary evolution in the $x+y$ and $x-y$ directions \cite{permute} and so a two dimensional version of the distribution of the walk on the line is obtained. This is shown in Fig.~\ref{fig:2dwalks}(a), where the initial state is taken as the walker at the origin with the separable coin state
\begin{equation}
\label{eqn:initcond1}
 \Psi (0,0,0)
 =  \frac{1}{2}\begin{pmatrix}1 \\i \\i\\ -1\end{pmatrix}.
\end{equation}
More interestingly, they consider the Grover coin:
\begin{equation}
G^4 = \frac{1}{2} \begin{pmatrix} -1 & 1 & 1 & 1\\ 1 & -1 & 1 & 1\\ 1 & 1 & -1 & 1\\ 1 & 1 & 1 & -1\end{pmatrix},
\end{equation}
and the discrete Fourier transform (DFT) coin:
\begin{equation} \label{eqn:dft4}
D^4 = \frac{1}{2} \begin{pmatrix}1 & 1 & 1 & 1\\ 1 & i & -1 & -i \\ 1 & -1 & 1 & -1\\ 1 & -i & -1 & i \end{pmatrix}.
\end{equation}
These operators were tested for a number of initial conditions. These coins are both unbiased, in that they distribute amplitude equally between each coin state. As for the walk on the line, they find the dynamics for a specific coin depend strongly on the choice of initial state.  However, the dynamics differ markedly depending on the coin used. The lowest and highest standard deviations obtained for the position of the walker were found using the Grover operator. It was observed that the reason for this is that regardless of the initial state, the distribution forms a central spike, with a ring around it which propagates outwards. The choice of initial condition controls the amount of amplitude that is situated in the central spike, and the amount of amplitude that is situated in the ring. The distribution for the DFT coin, given the initial state of Eq.~(\ref{eqn:initcond1}), is shown in  Fig.~\ref{fig:2dwalks}(b). The distribution for the Grover coin, where the initial state is taken to be the walker at the origin, with the coin state
\begin{equation}
\label{eqn:initcond2}
 \Psi (0,0,0)
 =  \frac{1}{2}\begin{pmatrix}1\\-1\\-1\\1\end{pmatrix},
\end{equation}
is shown in Fig.~\ref{fig:2dwalks}(c).
Additionally, the authors of \cite{Tregenna03} studied the set of unbiased four dimensional unitary operators with entries equal to either $\pm 1/2 $ or $\pm i/2$ which, when the leading diagonal entry is selected to be $1/2$, gives 640 unitary operators. These operators were found to produce ten different spreading rates, with the DFT, Hadamard and Grover all being different.

\section{Definition}
\label{sec:analytics}
\begin{figure}[th!] 
  \subfigure  
  {(a)\resizebox{0.8\columnwidth}{!}{\includegraphics[trim= 2cm 1.4cm 2cm 1.4cm]{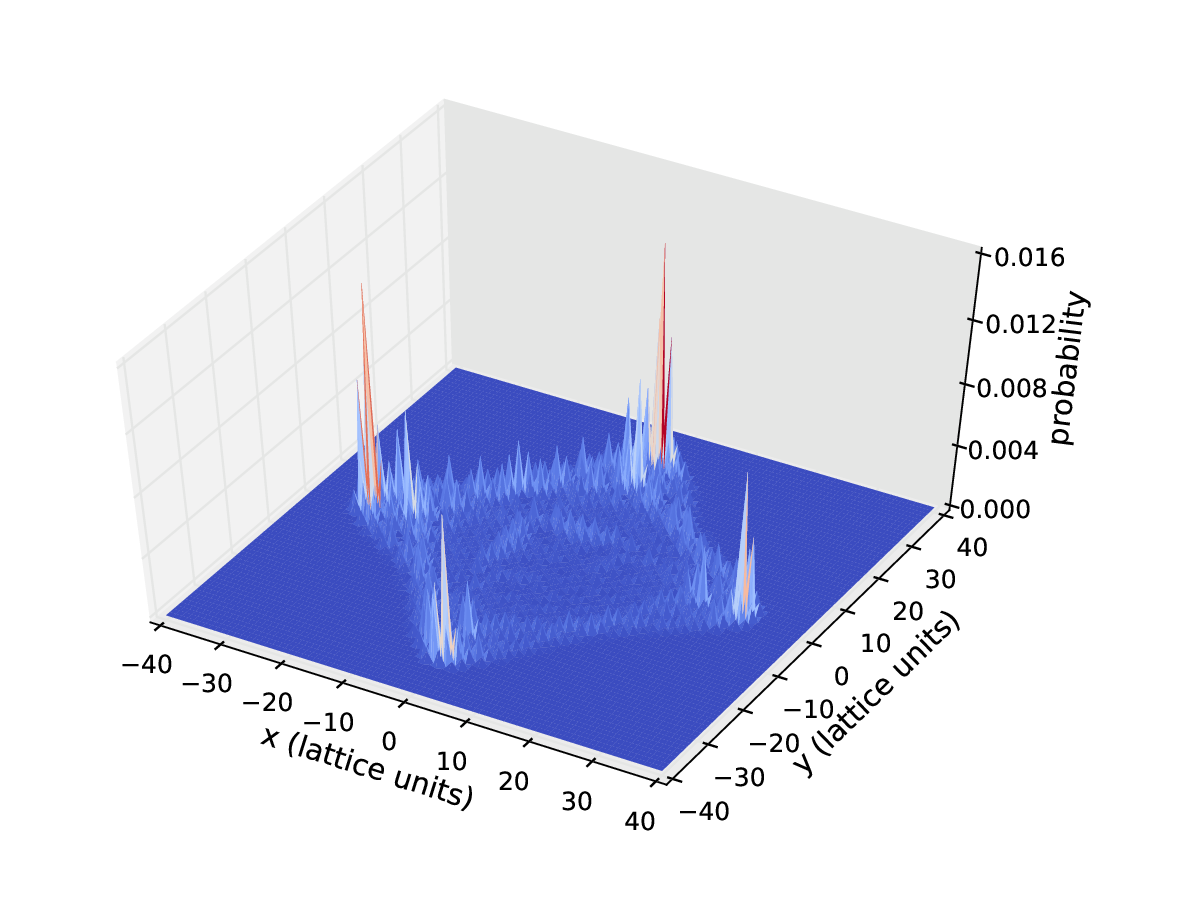}}} \\ 
  \subfigure  
  {(b)\resizebox{0.8\columnwidth}{!}{\includegraphics[trim= 2cm 1.4cm 2cm 1.4cm]{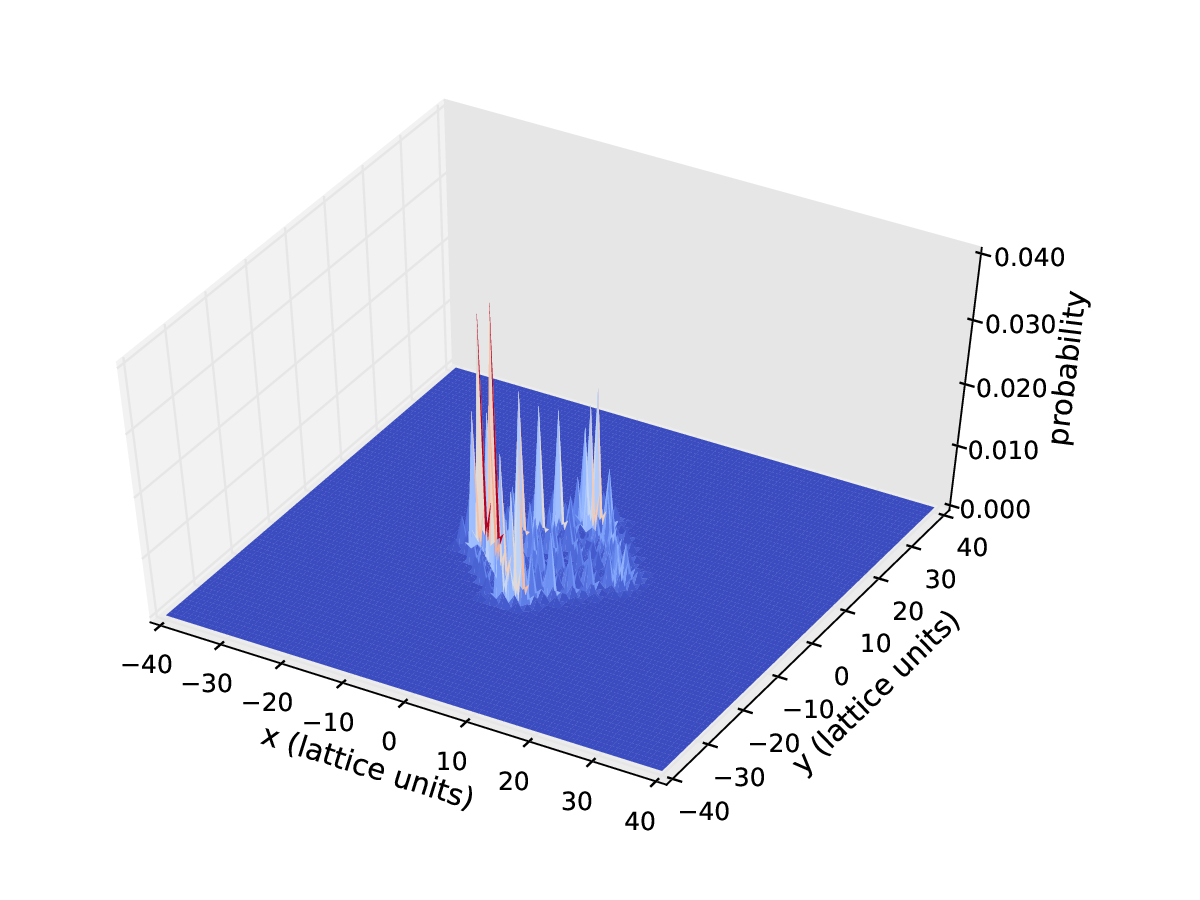}} } 
\caption{(Color online) Probability distributions arising after 100 steps of a) a typical nonreversal quantum walk and b) a typical nonrepeating quantum walk, with the initial conditions given by Eq.~(\ref{eqn:initcond1}) and Eq.~(\ref{eqn:initcond2}) respectively.}
\label{fig:sawalks}
\end{figure}

We now define the nonrepeating and nonreversal walks in terms of particular choices for the coin operator.  The first coin we will consider will generate the nonrepeating quantum walk and hence will be called the nonrepeating coin, denoted $C^{!rep}$. This coin is defined by:
\begin{widetext}
\begin{equation}
\label{eqn:sacoin}
C^{!rep}=\left(
\begin{array}{cccc}
0&\lambda e^{i\alpha}&\gamma e^{i\beta}&f(\lambda, \gamma)e^{i\theta} \\
\lambda e^{-i( \phi  + \delta + \alpha  )}&0&-f(\lambda, \gamma)e^{i(\psi -\theta +\beta )}&\gamma e^{i\psi}\\
-\gamma e^{-i( \delta +\alpha + \psi)}&-f(\lambda, \gamma)e^{i(\phi -\theta +\alpha )}&0&\lambda e^{i\phi }\\
f(\lambda, \gamma)e^{i(\theta - \alpha - \psi -\phi -\beta  )}&-\gamma e^{i(\delta+\alpha-\beta )}&\lambda e^{i\delta}&0
\end{array}
\right),
\end{equation} 
\end{widetext}
where all of the variables are real, $ 0 \leq \gamma^2 + \lambda^2 \leq 1,$  and   $f(\lambda, \gamma)=\sqrt{1-(\lambda^2+\gamma^2)}$.
This is the most general $SU(4)$ operator with zeros on the diagonal. 
It is clear, with reference to the shift operator defined in Eq.~(\ref{shift}), that the coin never permits amplitude to move in the same direction in two consecutive steps, and so it is natural to refer to this as a nonrepeating walk. We now define the nonreversal coin operator, in terms of a permutation of the nonrepeating coin operator, by:
\begin{equation}C^{!rev}=  C^{!rep} \cdot \left( \begin{array}{cccc} 0 & 0 & 0 & 1 \\ 0 & 0 & 1 & 0 \\ 0 & 1 & 0 & 0 \\ 1 & 0 & 0 & 0 \end{array} \right).\end{equation}
In analogy to the nonrepeating walk, the walk defined by this coin never permits amplitude to move back to the vertex where it was at the previous time step, and so hence is a nonreversal quantum walk. It is important to note that the interpretations of each of these walks is strongly linked to the definition of the shift operator. If an alternative definition is used, such as that in \cite{Kendon06}, then the interpretations of the walks created by these coins is changed. Whilst the nonreversal walker is a single particle, when interpreted as a dimer it is presumed that the two parts of the nonreversal walker are distinguishable, so one leads the other, as shown in Fig.~\ref{fig:dimer}.
\newline
\indent
As the walk dynamics that are obtained from the nonreversal and the nonrepeating coins have many properties in common we discuss both together. A simple example of a nonreversal coin, used to produce the probability distribution shown in Fig.~\ref{fig:sawalks}, takes $\theta=\phi= \frac{3\pi}{4}$, $\alpha = \beta = \delta = \psi = \frac{-\pi}{4}$ and $\lambda=\gamma=f(\lambda,\gamma)=\frac{1}{\sqrt{3}}$ in $C^{!rev}$. This leads to the following coin:
\begin{equation}
\label{eqn:sacoin2}
C^{1}=\frac{e^{-i \frac{\pi}{4}}}{\sqrt3}\left(
\begin{array}{cccc}
-1&1&1&0\\
1&1&0&1\\
-1&0&-1&1 \\
0&1&-1&-1
\end{array} \right),
\end{equation}
where the global phase factor can be dropped. Fig.~\ref{fig:sawalks} shows examples of typical probability distribution arising from a nonreversal and a nonrepeating quantum walk. The nonreversal walk displays a greater average radial distance from the origin than the nonrepeating walk. This is quantified in section~\ref{sec:numerics}.
In both cases, the dynamics are similar for all initial conditions, tracing out roughly a diamond shape, larger for the nonreversal walk, with peaks at each corner. The initial condition determines the height and number of distinctive peaks. In the case of the nonreversal walk, it is possible to see a smaller square insider the larger outer diamond that is characteristic of this walk. In the case of the nonrepeating walk, the outline of the possible sites that the walk can have reached after $t$ steps is given by a square with sides of length $t$ ($t+1$) if $t$ is even (odd) centered on the origin with the sides parallel to the $x$ and $y$ axes. However, it can be seen from Fig.~\ref{fig:sawalks}(b) that the characteristic shape of the peaks of the probability amplitude for this walk is also a diamond, as in
the nonreversal case, and with dimensions much smaller than $t$.

\section{Fourier analysis}
\label{sec:nr}

In order to analytically study the large $t$ behaviour of the quantum walks defined above, we use Fourier analysis. The Fourier transform from position space to momentum space is
\begin{equation} \hat{\Psi}(k_x, k_y, t) = \sum_{x,y} \Psi(x,y,t) e^{i(k_xx+k_yy)} ,\end{equation}
with the inverse transform given by
\begin{equation} \Psi(x, y, t) =\int_{-\pi}^{\pi}\int_{-\pi}^{\pi} \frac{dk_x dk_y}{(2 \pi)^2} \hat{\Psi}(k_x,k_y,t) e^{-i(k_xx+k_yy)}. \end{equation}
In momentum space the shift operator is given by
\begin{equation}S(k_x,k_y)= \left(
\begin{array}{cccc} e^{ik_x} & 0 & 0 & 0 \\ 0& e^{ik_y} & 0 & 0 \\  0& 0 & e^{-ik_y}  & 0 \\ 0&0& 0 & e^{-ik_x}   \end{array}\right).\end{equation}
The walk then evolves by the recurrence relation 
\begin{equation} \hat{\Psi}(k_x,k_y,t+1)=U_{c}(k_x,k_y)\hat{\Psi}(k_x,k_y,t), \end{equation} where 
$ U_{c}(k_x,k_y) =  S(k_x,k_y)C^{c}$ and $C^{c}$ is the chosen coin operator. Using the notation $\hat{\Psi}_t = \hat{\Psi}(k_x,k_y,t)$ and $\Psi_t =\Psi(x,y,t)$, we can rewrite this as  \begin{equation} \hat{\Psi}_t=U_{c}(k_x,k_y)^t \hat{\Psi}_0 \end{equation} 
As the walker is initially at the origin $\hat{\Psi}_0$ is constant in both $k_x$ and $k_y$. For analytical purposes, instead of considering moments in terms of the radial distance from the origin we consider the joint moments of the position operators in the $x$ and $y$ directions, denoted $X$ and $Y$ respectively. For a two dimensional quantum walk, these are given by 
\begin{multline} \left< X^{\xi}_t Y^{\chi}_t \right>_{\Psi}  = \sum _{x,y \in\mathbb{Z} }\Psi^{\dagger}_t  x^{\xi} y^{\chi} \Psi_t \\
 = \int_{-\pi}^{\pi}\int_{-\pi}^{\pi} \frac{dk_x dk_y}{(2 \pi)^2} \hat{\Psi}^{\dagger}_t \left(i \frac{\partial}{\partial k_x} \right)^{\xi} \left(i \frac{\partial}{\partial k_y} \right)^{\chi}  \hat{\Psi}_t, \label{moments} \end{multline}
where $i \frac{\partial}{\partial k_x}$ and  $i\frac{\partial}{\partial k_y}$ are the momentum space representations of the position operators $X$ and $Y$. In order to calculate the state of the walker at time $t$ for a particular $C^c$ we need to calculate the eigensystem of $U_c$. We will first of all take $U_{!rep}$ and show that in this case the even moments, i.e., when $\xi+\chi$ is even, are independent of the initial state of the walker for large $t$. In Appendix A it is shown that the eigenvalues of $ U_{!rep}(k_x,k_y)$, denoted by $p_j$, can be expressed as 
\begin{equation} p_1=-p_2=p_3^*=-p_4^*=e^{i \omega(k_x,k_y)}, \end{equation}
where $\omega$ is a function of $k_x$, $k_y$ and all 8 coin parameters. This is because the characteristic equation of $U_{!rep}(k_x,k_y)$ is of the
form
\begin{equation} p^4 + A p^2 + B = 0, \label{pchareq} \end{equation}
where $A$ and $B$ are real.  We label a corresponding set of orthonormal eigenvectors by $\ket{v_j(k_x,k_y)}$, $j=1,2,3,4$. We will drop the $k_x$ and $k_y$ dependence from the notation. We may represent the state in terms of the eigensystem by:
\begin{equation}  \hat{\Psi}_t=  U(k_x,k_y)^t \hat{\Psi}_0 = \sum_{j=1}^4p_j^t\expect{v_j |\Psi_0}\ket{v_j}.\end{equation}
We will now show that the joint moments, as defined in Eq.~(\ref{moments}), are asymptotically independent of $\Psi_0$ using the method of Grimmett \emph{et al.} \cite{Grimmett04,Watabe08} to calculate the large $t$ expression for the moments. First consider

\begin{widetext}
\begin{multline}
\left(i \frac{\partial}{\partial k_x} \right)^{\xi} \left(i \frac{\partial}{\partial k_y} \right)^{\chi}  \hat{\Psi}_t = 
\left(i \frac{\partial}{\partial k_x} \right)^{\xi}  \left(i \frac{\partial}{\partial k_y} \right)^{\chi}   \sum_{j=1}^2 (-1)^{(j-1)t}\left(
e^{i \omega t}\expect{v_j|\Psi_0} \ket{v_j} + e^{-i\omega t}\expect{v_{j+2}|\Psi_0} \ket{v_{j+2}} \right)
\\
= \Big( (-1)^{\xi + \chi}e^{i \omega t}\sum_{j=1}^2 (-1)^{(j-1)t}\expect{v_j|\Psi_0} \ket{v_j} 
+  e^{-i \omega t}\sum_{j=3}^4 (-1)^{(j-3)t}\expect{v_j|\Psi_0} \ket{v_j}  \Big) t^{\xi + \chi} \left( \frac{\partial  \omega}{\partial k_x} \right)^{\xi} \left( \frac{\partial  \omega}{\partial k_y} \right)^{\chi} + \mathcal{O}(t^{\xi +\chi -1}).
\end{multline}
Considering the whole of the integrand in Eq.~(\ref{moments}), we then have that
\begin{multline}
\hat{\Psi}^{\dagger}_t \left(i \frac{\partial}{\partial k_x} \right)^{\xi} \left(i \frac{\partial}{\partial k_y} \right)^{\chi}  \hat{\Psi}_t 
=\Big\{(-1)^{\xi+\chi} \sum_{j=1}^2 \left|\expect{v_j|\Psi_0}\right|^2+\sum_{j=3}^{4} \left|\expect{v_j|\Psi_0} \right|^2 \Big\}
t^{\xi+\chi} \left( \frac{\partial  \omega }{\partial k_x} \right)^{\xi} \left( \frac{\partial  \omega }{\partial k_y} \right)^{\chi} + \mathcal{O}(t^{\xi +\chi -1}).
\end{multline}
\end{widetext}

Now as $\sum_{j=1}^4 |\langle v_j|\Psi_0 \rangle|^2=1$ then if $\xi+\chi=2n$, $n \in \mathbb{N}$ we have
\begin{equation}\hat{\Psi}^{\dagger}_t X^{\xi} Y^{\chi}  \hat{\Psi}_t =t^{\xi+\chi} \left( \frac{\partial  \omega }{\partial k_x} \right)^{\xi} \left( \frac{\partial  \omega }{\partial k_y} \right)^{\chi} + \mathcal{O}(t^{\xi +\chi -1}) \label{eqn:eqn}. \end{equation}

Therefore, by substituting Eq.~(\ref{eqn:eqn}) into Eq.~(\ref{moments}) we see that under the condition $\xi+\chi=2n$, and in the asymptotic limit of large $t$, the moments are independent of the initial state of the walker and are a function of the coin parameters only. From appendix A it can be seen that the moments (asymptotically) depend only on the five parameters
$m_1 = \alpha - \beta + \delta  + \psi$, $m_2=\phi+\delta$, $m_3=\phi + \alpha - 2 \theta  + \psi + \beta $, $\lambda$ and $\gamma$. This is because $\omega(k_x, k_y)$ can be written as a function of $k_x,$ $k_y,$ $m_1,$ $m_2,$ $m_3,$ $\lambda$ and $\gamma$. Although this is an asymptotic proof, numerical results show that this is true for any $t$, suggesting that the dependence on the initial states cancels in a similar way for all orders, not just the leading order. A similar result also holds for the Hadamard walk on a lattice, as defined by the coin of Eq.~(\ref{Hadcoin}) and the shift of Eq.~(\ref{shift}). For this separable coin \cite{permute}, in the limit of large $t$, $\langle (X_t+Y_t)^{\xi},(X_t-Y_t)^{\chi} \rangle$ is independent of the initial state when \emph{both} $\xi$ and $\chi$ are even. This follows directly from the properties of a Hadamard walk on a line and is shown briefly in appendix C.

\begin{figure}[th!]  
  \subfigure 
  {(a)\resizebox{0.85\columnwidth}{!}{\includegraphics{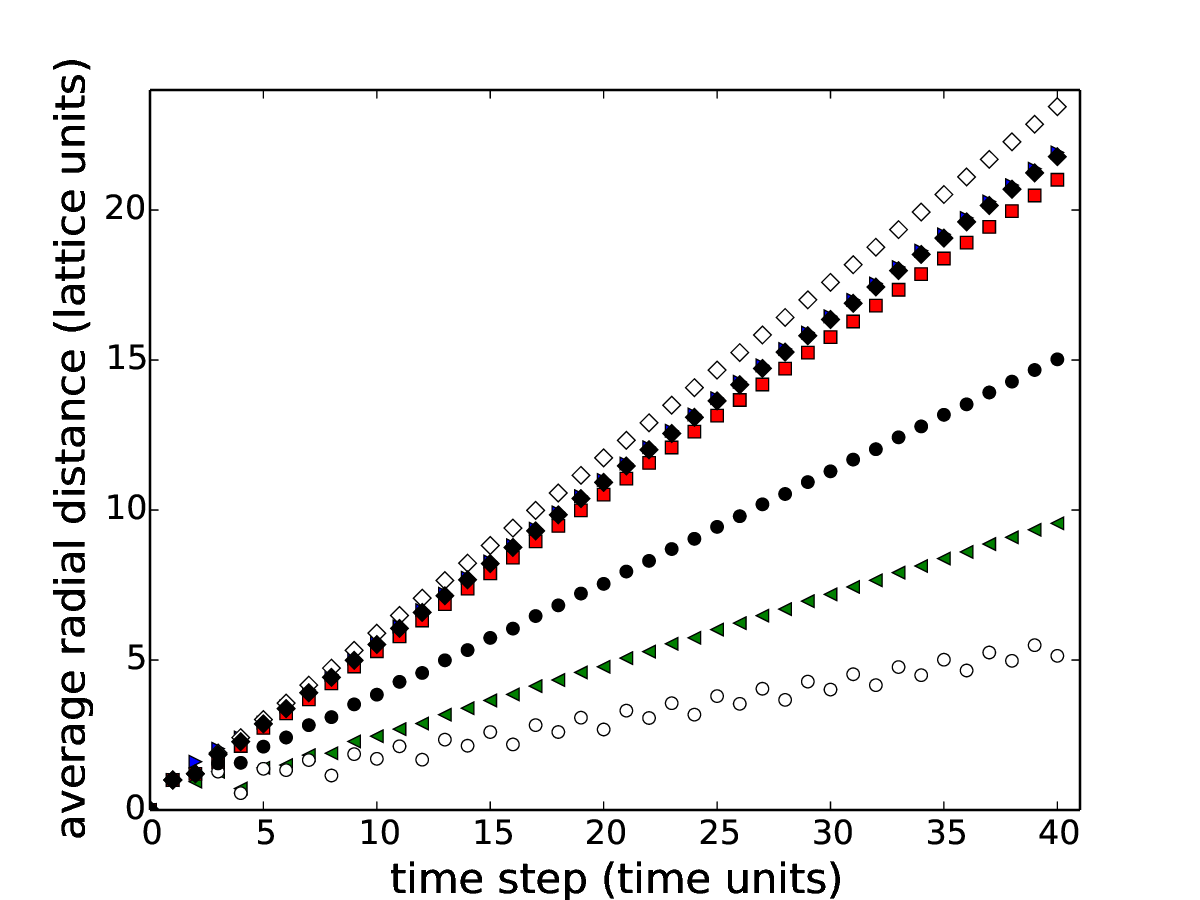}}} \\ 
  \subfigure 
  {(b)\resizebox{0.85\columnwidth}{!}{\includegraphics{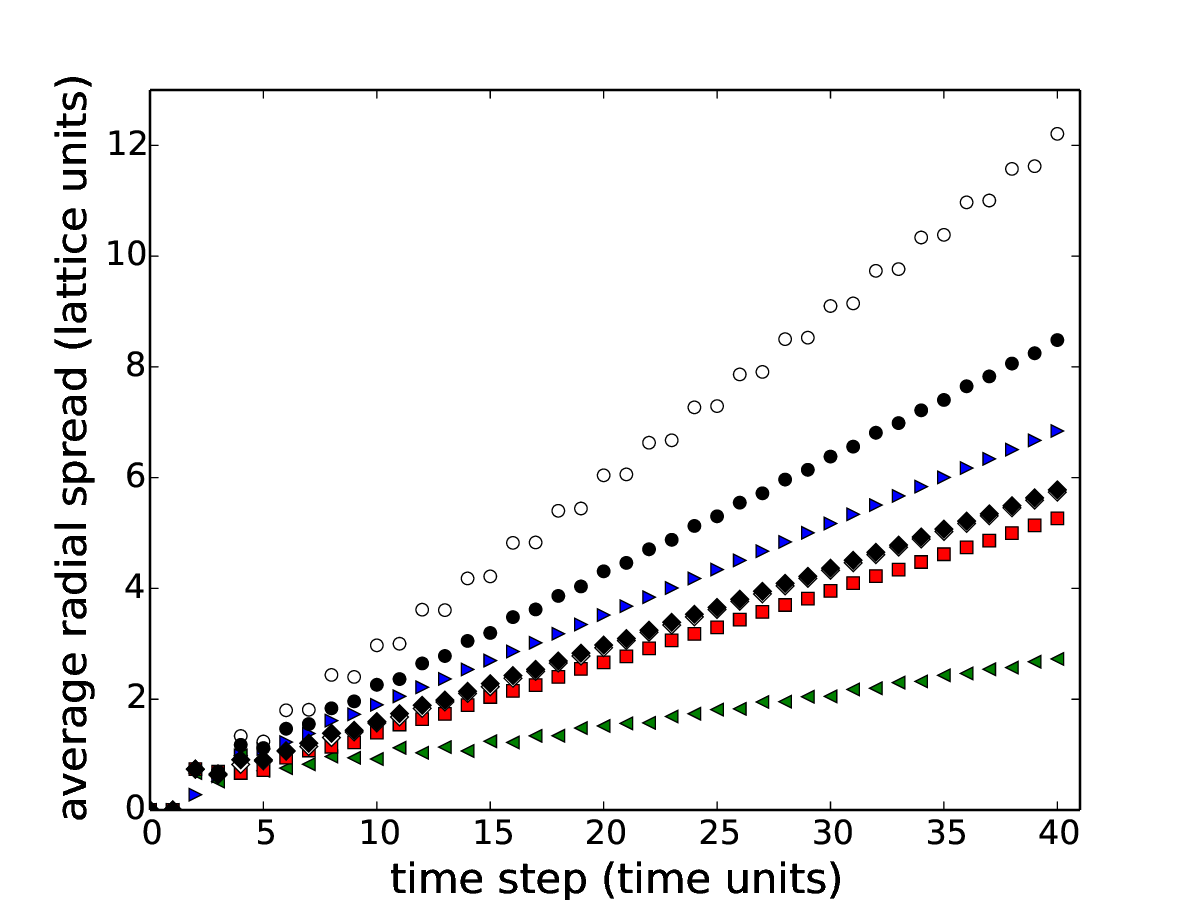}}} 
\caption{(Color online) Comparison of the nonreversal (blue right triangles), nonrepeating (green left triangles), 
Hadamard (red squares), Grover (open diamonds and circles) and DFT (black diamonds and circles) coin operators in 
terms of a) the average radial distance from the origin, defined in Eq.~(\ref{eqn:mean}) and b) the standard deviation of the radial distance, defined in Eq.~(\ref{eqn:sd}). 
The diamonds are the largest and the circles the smallest average radial distances for different initial states for those two coins; the
values are the same for all initial states for the other three coins.} 
\label{fig:compare}

\end{figure}

\section{The nonreversal walk}
\label{sec:numerics}

The result derived above applies only to walks using the operator $U_{!rep}$. We conjecture that the same result holds for the nonreversal walk. As its eigensystem is not tractable using the same methods as for the nonrepeating walk, as shown in Appendix B, the nonreversal walk is treated numerically rather than analytically. These walks were investigated by varying the parameters in the coin as well as the initial condition. 
Independent uniformly random choices for each variable were used to generate 500 coins, and the walks arising from these operators were investigated using roughly 1000 initial conditions for the coin state of a walker at the origin, parameterised using
\begin{equation}
\label{eqn:initstate}
 \Psi (0,0,0)
 =  \begin{pmatrix} \cos\theta_3 \\ 
 					e^{i\phi_3}\sin\theta_3\cos\theta_2 \\ 
 					e^{i\phi_2}\sin\theta_3\sin\theta_2\cos\theta_1 \\ 
 					e^{i\phi_1}\sin\theta_3\sin\theta_2\sin\theta_1 
 	\end{pmatrix},
\end{equation}
i.e., uniformly according to the Haar measure \cite{Nemoto00,rowe99}. This means that we sample $0<\phi_i \leq 1$ and $0 < \cos \theta_i \leq 1$, for $i=1,2,3$ uniformly.  It was found numerically that, for all choices of initial condition, the mean radial distance from the origin, as given by Eq.~(\ref{eqn:mean}), and the standard deviation, given by Eq.~(\ref{eqn:sd}), of the walker are constant at a given time $t$. Further investigations suggest that all the joint moments of the distribution where the $x$ and $y$ exponents sum to an even number are independent of the initial condition. 

To compare with the nonrepeating coin, we tested whether the moments were constant if the five parameters $m_1 = \alpha - \beta + \delta  + \psi$, $m_2=\phi+\delta$, $m_3=\phi + \alpha - 2 \theta  + \psi + \beta $, $\lambda$ and $\gamma$ were held constant whilst varying their constituents. As in this case only twenty coins were tested, the results are not conclusive, but it appears that these same five parameters determine the moments in the case of the nonreversal coin. 

The properties of both walks contrast strongly with those arising from previously studied non-separable coins. The mean and standard deviations as a function of time are shown for a variety of coins in Fig.~\ref{fig:compare}. For the nonreversal, nonrepeating and Hadamard walks these are independent of the initial condition. For the Grover and DFT walks this is not the case and so the initial conditions which give the largest and smallest values for $\expect{r}$ are plotted for both coins. Fig.~\ref{fig:compare} shows that both the average radial distance and the average radial spread are greater for the nonreversal than for the nonrepeating walk. The average radial distance for both of these walks lies within the range of the possible values achievable with the Grover coin with the use of specific initial coin states.

\section{Conclusion}
\label{sec:conclusion}

We have introduced two previously unstudied types of coin operator for the discrete time quantum walk, the nonrepeating and nonreversal coins. We have shown that they have some notable properties, namely that the mean and standard deviation of the radial distance from the origin of the walker is independent of the choice of initial condition, in contrast to all the commonly used non-separable coin operators. The standard deviation still grows linearly with $t$ for much the same reason as it does for the walk on the line, as the coin operator always ensures that some amplitude moves away from the starting point with each step. We have shown, analytically for the nonrepeating operator and numerically for the nonreversal operator, that the \emph{even} joint moments of the $x$ and $y$ positions are independent of the initial condition of the walker; the odd moments do depend on the initial condition. The even moments of the nonrepeating walk depend on five parameters derived from the nonrepeating coin. We have also provided numerical evidence that the moments of the nonreversal walk depend on the same five parameters. 

For future work, it would be interesting to investigate the properties of the nonrepeating and nonreversal walks on other lattices besides the square lattice.    The self-avoiding random walk has been shown to have macroscopic properties which are independent of the choice of lattice.  In order to see if this property carries over into the quantum case, analogous coins of varying dimension are required, which can be constructed by parameterizing a unitary matrix with the appropriate pattern of zero entries.  Further coins that exhibit the same analytical properties as the
nonrepeating coin, i.e., the even moments are independent of the
initial state, could be constructed by creating coin operators with
eigenvalues that are the solutions to characteristic equations that have the
form of Eq.~(\ref{pchareq}).

\begin{acknowledgments}
T.J.P. was funded by a University of Leeds Research Scholarship. K.E.B. was funded by the UK Engineering and Physical Sciences Research Council. V.M.K. was funded by a UK Royal Society Research Fellowship. T.J.P., B.H., and A.B. received support through a scholarship from the UK Royal Society and the UK Engineering and Physical Sciences Research Council. V.P. was supported by the International Association for the Exchange of Students for Technical Experience and Ministry of Education, Science and Technological Development of the Republic of Serbia (project ON171025). S.M. was funded by JFT Grants No. 15619 and No. ANR-10-JCJC-0208 CausaQ.
\end{acknowledgments}

\bibliography{MyLibrary}

\begin{thebibliography}{28}
\bibitem [{per()}]{permute}%
  \BibitemOpen
  \href@noop {} {}\bibinfo {note} {In \cite{Tregenna03,Mackay02} the quantum
  walk takes place on the diagonals with the shift operator defined in the
  orthogonal $x+y$ and $x-y$ directions (i.e the shifts are by one lattice
  point in both $x$ and $y$), with the coin basis given by the tensor product
  $(+,-)_y \otimes (+,-)_x = (++,+-,-+,--)$. This is a $45^{\circ}$ rotation
  from the coin basis used here of $(x_+,y_+,y_-,x_-)$. Equivalent results to
  \cite{Tregenna03,Mackay02} hold for the shift operator as defined here which
  can be seen using the change of variables in the (momentum space) shift
  operator of $k_{\pm} = \frac{1}{2}(k_x \pm k_y)$, as is demonstrated in
  appendix \ref{Hadap}.}\BibitemShut {Stop}%
\end{thebibliography}

\appendix

\section{}
Here we calculate the eigenvalues of $U_{!rep}(k_x,k_y)$.
Using algebraic manipulation software it can be shown that the characteristic equation for this matrix is given by                                                 
\[ p^4+2p^2\left(\gamma^2\cos\Theta_1 -\lambda^2\cos\Theta_2 - f(\lambda,\gamma)^2 \cos \Theta_3  \right)+1=0,\]
where $\Theta_1=m_1- k_x + k_y $, $\Theta_2=m_2 -k_x -k_y$ and $\Theta_3=m_3$ with $m_1=\alpha - \beta + \delta +\psi$, $m_2=\phi+\delta$ and $m_3 = \phi + \alpha - 2 \theta + \psi + \beta$.
If we solve this equation for $p$ we will obtain the 4 eigenvalues. As there are no first or third order terms in the characteristic equation, if $p$ is a solution then so to is $-p$. As the coefficients are real then if $p$ is a solution then so is $p^*$. 
As the coin is unitary we therefore have that the eigenvalues can be written as $p_1=e^{i \omega(k_x,k_y)}$, $p_2=-e^{i \omega(k_x,k_y)}$, $p_3=e^{-i \omega(k_x,k_y)}$, $p_4=-e^{-i \omega(k_x,k_y)}$ 
where $\omega$ is a function of $\lambda$, $\gamma$, $\Theta_1$, $\Theta_2$, and $\Theta_3$ and so hence is a function of $k_x$, $k_y$ and the coin parameters $\lambda$, $\gamma$, $m_1$, $m_2$ and $m_3$. By solving the effective quadratic it can be shown that the solutions are given by $p_i=\pm_s \sqrt{a_r \pm_t i a_i }$, where the subscripts denote that $\pm_s$ and $\pm_t$ are independent and  $a_r= - b$, $a_i=\left| \sqrt{b^2-1} \right| $, $b=\gamma^2\cos\Theta_1 -\lambda^2\cos\Theta_2 - f(\lambda,\gamma)^2\cos \Theta_3$.

\section{}
The characteristic equation for $U_{!rev}$ can be shown to be
\[p^4+ \Delta p^3+ \Xi p^2 + \Delta^* p +1 = 0, \]
where $\Delta =  f(\lambda,\gamma) ( e^{i(b_1-k_y)} +e^{i(b_2+k_y)}- e^{-i(b_1+b_2+\theta+k_x)}- e^{i(\theta+k_x)} ) $ and $\Xi= 2 ( f(\lambda,\gamma)^2\cos(b_1+b_2) +(\lambda^2-1)\cos(k_x+k_y +b_2+\theta) +(\gamma^2-1)\cos(k_x-k_y +b_1 +\theta) ) $ with $b_1=\alpha+\phi-\theta$ and $b_2=\beta+\psi-\theta$. This quartic does not in general have the properties of that in appendix A. However if $b_1=-b_2$ then $\Delta=\Delta^*$ and so the characteristic equation is quasi-symmetric and has real parameters. As the parameters are real, if $p$ is a solution then so is $p^{*}$. We can then write the solutions in terms of two parameters $\omega_1$ and $\omega_2$ such that $p_1=p_2^{*}=e^{i \omega_1}$ and $p_3=p_4^{*}=e^{i \omega_2}$. It can be shown that the solutions are of the form
\begin{multline*}
p =  - \frac{\Delta}{2} \pm_s \frac{\sqrt{\Delta^2-4(\Xi-2)}}{2} 
 \\ \pm_t   \frac{\sqrt{2\Delta^2-4(\Xi+2)\mp_s \Delta\sqrt{\Delta^2-4(\Xi-2)}}}{4}.
\end{multline*}
We can therefore show that we cannot write the solutions in the form derived in Appendix A, that is we cannot write $p_1=p_2^{*}=-p_3=-p_4^{*}=e^{i \omega}$ and so the proof method for $U_{!rep}$ does not follow for $U_{!rev}$.

\section{}
\label{Hadap}
Here we show that the Hadamard walk has the property that $\langle (X_t+Y_t)^{\xi}  (X_t-Y_t)^{\chi}  \rangle$ is independent of the initial state of a walker at the origin if $\xi$ and $\chi$ are even, in the limit of large $t$.
The Hadamard walk on the lattice is given by the coin $C^{H}=H\otimes H$ of Eq.~(\ref{Hadcoin}). The walk then evolves in momentum space via the unitary operator \[U_H = S(k_x,k_y)C^{H}=S_{k_+} H\otimes S_{k_-} H,\] where $k_{\pm}=\frac{1}{2}(k_x\pm k_y )$ and
\[ S_{k_{\pm}} = \begin{pmatrix} e^{i k_{\pm}} & 0 \\ 0 & e^{-i k_{\pm}} \end{pmatrix} .\]
It is straightforward to show that the eigenvalues of $S_{k_{\pm}}H$ can be written as $e_1(k_{\pm})=-e_2^{*}(k_{\pm})=e^{i \omega (k_{\pm})}$ and hence the eigenvalues of $U_H$ can be written as $p_1=p_4^{*}=e^{i(\omega(k_+) + \omega(k_-))}$ and $p_2=p_3^{*}=-e^{i(\omega(k_+) - \omega(k_-))}$.
We note that
\[ i\left(  \frac{\partial}{\partial k_x} \pm\frac{\partial}{\partial k_y} \right)(\omega(k_+) + \omega(k_-)) = i\frac{d \omega(k_{\pm} )}{d k_{\pm}}   .\]
Hence, following the same method as used for proving the initial state independence of the moments for the nonrepeating coin, it is possible to show that in the limit of large $t$, and when both $\chi$ and $\xi$ are even, $\langle (X_t+Y_t)^{\xi}  (X_t-Y_t)^{\chi}  \rangle$ is independent of the initial state of a walker at the origin for the $U_H$ quantum walk.
\end{document}